\documentclass[letterpaper,12pt]{article}
\usepackage{tabularx} % extra features for tabular environment
\usepackage{amsmath}  % improve math presentation
\usepackage{graphicx} % takes care of graphic including machinery
\usepackage[margin=1in,letterpaper]{geometry} % decreases margins
\usepackage{epstopdf}
\usepackage{cite} % takes care of citations
\usepackage[final]{hyperref} % adds hyper links inside the generated pdf file
\hypersetup{
	colorlinks=true,       % false: boxed links; true: colored links
	linkcolor=blue,        % color of internal links
	citecolor=blue,        % color of links to bibliography
	filecolor=magenta,     % color of file links
	urlcolor=blue         
}
\begin{document}

\title{\textbf{Estimating the D-line Energy Splitting of Alkali Metals Using a Relativistically Corrected Screening Constant}
\author{E. Holt Stewart, Douglas A. Barlow \footnote{dabarlow@sewanee.edu}~ \footnotesize{ORCID: 0000-0003-0117-5139}\\
\small Department of Physics, The University of the South, Sewanee TN, 37383 USA\\}}

\date{}
\maketitle

\begin{abstract}
We report here on how a known method from standard perturbation theory for estimating the energy of the D-line splitting in hydrogen can be
modified to effectively approximate this quantity
for all of the alkali metals. The approach utilizes a
Rayleigh-Schrödinger perturbation theory first
order correction to the energy. The perturbing
Hamiltonian is the standard relativistically corrected
spin-orbit Hamiltonian. From this, one derives an
energy difference between the doublet lines that is
theoretically appropriate for any one electron atom.
This energy difference is written in terms of the
Bohr energy. The results are good for hydrogen
but, as expected, are significantly off from the
experimental values for the multi-electron alkali
metals. We show here that by replacing the Bohr
energy with a first ionization potential, greatly
improved values for the D-line splitting energy
can be obtained from the model. However, this
approach overestimates the splitting energy for
the light alkali metals and underestimates it for
the heavier ones. The best result was for Rb where
the estimate only varied from the experimentally
reported value by 3.2\%. Accurate results are
finally obtained from the model for all of the
alkali metals when the original Bohr energy is
adjusted with an appropriate screening constant.
Screening constants generated using the Slater
scheme however, which yield accurate estimates
for ionization potentials, do not give the correct results for the D-line splitting energies. By fitting
our results to reported energy splitting values we
find that the correct screening constant is a function
of the atomic number. We show that this result
can be reproduced via a simple model with a
relativistic correction for the contraction of the
radius of the inner shell electrons.
\end{abstract}


\clearpage

\section{Introduction}

The principle line spectra of an excited hydrogen gas
sample were described empirically by J. J. Balmer
in 1885. However, soon thereafter, an additional
splitting in some of these principle lines was
observed, an effect referred to as \textit{fine structure splitting}. About a decade later, Michelson and
Morley discovered fine structure splitting in the
spectrum of sodium \cite{Michelson}. By the early 20$^{\text{th}}$ century,
the Balmer line spectra of hydrogen had been
described theoretically by the Bohr formula where
lines were taken to be the result of energy
transitions between quantum states with principle
quantum number $n =2$ and $n = 3, 4, 5, \cdots$.

After the appearance of Bohr’s formula, a
quantitative explanation for fine structure in
hydrogen spectra would soon be developed. Since
that time, a variety of complex splittings have
been observed in excited gas spectra but we will
restrict our discussion here to a particular class of
fine structure. This type of splitting occurs when one principle line is separated into two closely
spaced bands, the so called doublets, or D-lines.

After the discovery of electron spin, a model for
coupling between spin and orbital angular momentum
was developed to explain the origin of the doublet
splitting in hydrogen. This theory was further
refined in 1927 by L. H. Thomas who included
relativistic effects thus leading to the mature result
that is still widely used today.

Unfortunately, since state functions are only fully
known for the hydrogen atom, the quantum spin-
orbit theory is only strictly appropriate for the
description of D-line splitting in hydrogen or a
one electron cation. However, this type of fine
structure is known to appear in the spectra of
many other elemental gases, most notably the famous
yellow doublet lines that appear in the spectra of
Na. This process is depicted schematically in
Figure 1.

\begin{figure}[h]
  \centering
  %\label{fig:7}
  \includegraphics[width = 7.0 cm]{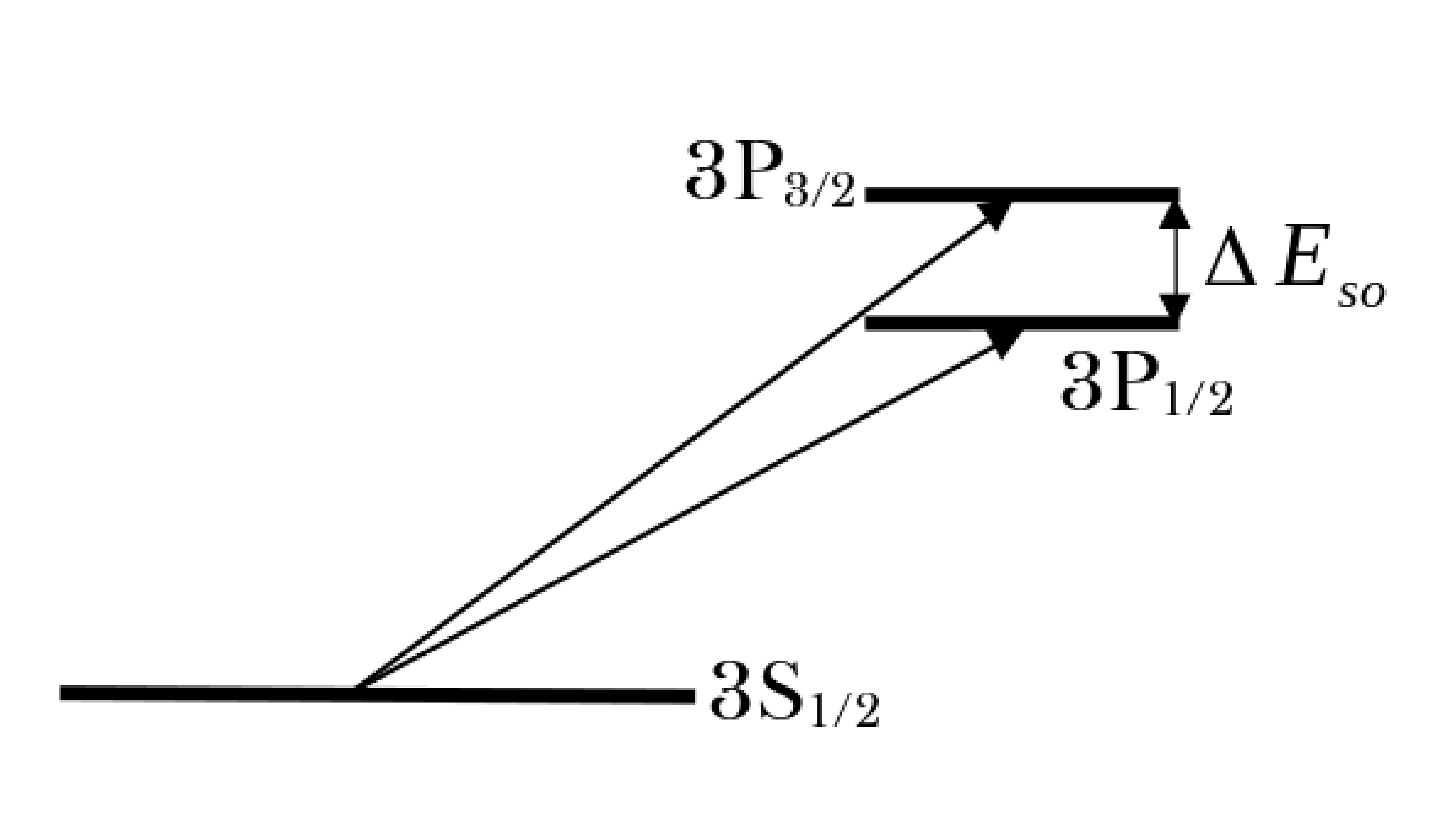}
  \caption{Schematic energy level diagram showing the two transitions that yield the yellow sodium doublet.}
  %\label{fig1}
\end{figure}

In Figure 1 the spectroscopic notation $nl_j$, is used
where $n$ is the principle quantum number, $l$ the
angular momentum quantum number where $l = S, P, D, \cdots$ and $j$ is the total angular momentum
number and $j = l \pm 1/2$. In fact, all of the alkali
metals display a very similar type of fine
structure. We label the energy difference between
the bands as $\Delta E_{so}$. A well-known method for
estimating $\Delta E_{so}$ for hydrogen is through Rayleigh-Schr\"{o}dinger perturbation theory where the spin-
orbit effect is accounted for by using a perturbing
spin-orbit Hamiltonian. The spin-orbit Hamiltonian
can be derived from classical considerations and
adjusted to include relativistic effects. Then,
through the perturbation correction to the energy to first order, a closed form expression can be
obtained for $\Delta E_{so}$. However, this result is only
valid for the one electron atom and as such predicts
the experimental splitting energy reasonably well
for hydrogen but fails to accurately predict the
same quantity for the heavier alkali metals.

When taking this approach, the final expression
for $\Delta E_{so}$ can be manipulated so that it is written in
terms of the classic Bohr energy $E_n$ , the fine
structure constant $\alpha$, the atomic number $Z$ and the
principle quantum number $n$. We show here that
by replacing this Bohr energy with the first
ionization potential of the element involved, this
expression for $\Delta E_{so}$ yields values for the D-level
splitting energies that are in better agreement with
experimental results for all of the first column
elements from Li to Fr. The best result was for Rb
where the estimate for $\Delta E_{so}$ was within 3.2\% of
the experimentally reported value.

We then show that this model can be improved,
and used to compute an accurate value for the D-
line splitting energy, if the Bohr energy is adjusted
with an appropriate screening constant. The well-
known Slater scheme for arriving at such a constant
was considered but screening constants computed
using this method did not yield accurate results.
From an empirical fit, it is shown that the correct
screening constant is a function of the atomic
number. We then demonstrate how a simple
screening constant model, involving a relativistic
correction for inner shell contraction, can be
developed. The final result has the screening
constant as a function of the atomic number with
constant coefficients that can be related to the
ratios of cationic to atomic radii.

The required ionization potentials were taken from
experimental reports and computed using the General
Atomic and Molecular Electronic Structure System
(GAMESS) \cite{GAMESS} at the density functional theory
(DFT) level. All data and parameter values used
to generate results mentioned in this report are
listed in tables.

In the next section we briefly outline Rayleigh-Schr\"{o}dinger
 first order perturbation theory and
then, in greater detail, review the derivation of the
spin-orbit Hamiltonian.

\section{Theory}

Rayleigh-Schr\"{o}dinger perturbation theory can be
used to obtain approximate solutions to a Schr\"{o}dinger
equation for cases where the differential equation
cannot be resolved analytically by known techniques.
To utilize this approach, one typically divides the
Hamiltonian into two terms. One term, the
unperturbed Hamiltonian, for which the analytic
solution to the Schr\"{o}dinger equation is known
and, the other, the perturbing Hamiltonian, which
then makes the system intractable. It is however a
requirement that the effects of the perturbing
Hamiltonian be small when compared to the non-
perturbing part.

So, to deal with spin-orbit splitting in the hydrogen
atom as a perturbation, the new Hamiltonian is
then written in terms of the unperturbed, non-
relativistic Hamiltonian for the hydrogen atom, $\hat{H}^o$,
, and a new perturbed Hamiltonian, $\hat{H}'$,
, multiplied by a smallness factor $\lambda$. That is
\begin{equation}
\hat{H} = \hat{H}^o + \lambda \hat{H}'~.
\label{eq:1.1}
\end{equation}

This method yields an expression for the new
system energy eigenvalues, $E'_n$,
that are shifted
somewhat due to the perturbation. The complete
derivation for this result has been covered in
detail previously and will not be fully reviewed
here \cite{Morrison, Razi}. The final result to first order is
\begin{equation} 
E'_n = E_n + \langle\Psi_n^o|\hat{H}_{so}|\Psi_n^o\rangle~,
\label{eq:1.2}
\end{equation}
where $E_n$ are the unperturbed Bohr energies, $\Psi_n^o$ the eigenfunctions of the unperturbed Hamiltonian and $\hat{H}_{so}$ the spin-orbit Hamiltonian where $\hat{H}_{so} = \lambda \hat{H}'$.

We now require an expression for the spin-orbit Hamiltonian, $\hat{H}_{so}$. This begins by writing the potential energy for an electron orbiting a nucleus of $Z$ protons. Letting the rest frame be at the electron, the electron sees an orbiting nucleus of charge $Ze$ where $e$ is fundamental charge. Therefore, there is a magnetic field of magnitude $B$ at the location of the electron. From classical electrodynamics, $\hat{H}_{so}$, is then just the potential of the electron with magnetic moment $\vec{\mu}$ in the field of the nucleus given as
\begin{equation}
\hat{H}_{so}= -\vec{\mu}\cdot \vec{B}~.
\label{eq:1.3}
\end{equation}

Using the Biot-Savart law one can write $\vec{B}$
terms of the current, as seen by the electron, and
then this can be further written in terms of the
orbital angular momentum, $\vec{L}$. The magnetic dipole
moment of the electron is then given in terms of
its spin angular momentum $\vec{S}$. However, since the
rest frame of the electron is not an inertial rest
frame, we must add a factor of $1/2$ to account for
Thomas procession so that
\begin{equation} 
H_{so} = \frac{Z e^2}{8\pi\epsilon_0}\left(\frac{1}{m^2c^2r^3}\right)\vec{S}\cdot\vec{L}~.
\label{eq:1.4}
\end{equation}

This spin-orbit Hamiltonian involves the vectors $\vec{S}$ and $\vec{L}$ but we require an operator form. Therefore, we create a new vector, $\vec{J}$, called the total angular momentum where
\begin{equation} 
\vec{J} = \vec{S} + \vec{L}~.
\label{eq:1.5}
\end{equation}
Using this relationship for the total angular momentum we can arrive at an operator form for $\vec{S}\cdot\vec{L}$. By evaluating the square of $\vec{S} + \vec{L}$, re-arranging and using Eq. (\ref{eq:1.5}), we arrive at
\begin{equation}
\vec{S}\cdot \vec{L} = \frac{1}{2}[\hat{J}^2-\hat{L}^2-\hat{S}^2]~.
\label{eq:1.6}
\end{equation}
Using Eq. (\ref{eq:1.6}) in Eq. (\ref{eq:1.4}) the spin-orbit Hamiltonian can be written as:
\begin{equation} 
H_{so} = \frac{Z e^2}{16\pi\epsilon_0}\left(\frac{1}{m^2c^2r^3}\right)[\hat{J}^2-\hat{L}^2-\hat{S}^2]~.
\label{eq:1.7}
\end{equation}
Letting the first order correction energy be $E_n^1 = \langle\Psi_n^o|\hat{H}_{so}|\Psi_n^o\rangle$ and then using $\hat{H}_{so}$ from Eq. (\ref{eq:1.7}) yields
\begin{equation}
E_n^1 = \frac{Z e^2}{16\pi\epsilon_0}\left(\frac{1}{m^2c^2}\right)\langle\psi_n | \frac{\hat{J}^2-\hat{L}^2-\hat{S}^2}{r^3} | \psi_n \rangle~.
\label{eq:1.8}
\end{equation}
The required integration over $r^{-3}$ can be accomplished using a recursion formula and a generating function for the associated Laguerre polynomials \cite{Johnson}. The eigenvalues are known for the operators $\hat{J}^2$, $\hat{L}^2$, and $\hat{S}^2$. Finally, after some rearranging, we can write the final result in terms of the Bohr energy and the fine structure constant $\alpha \approx 1/137$.
\begin{equation} 
E_n^1 = \frac{Z^2|E_n|\alpha^2}{2n}\frac{[j(j+1)-l(l+1)-\frac{3}{4}]}{l(l+\frac{1}{2})(l+1)}~.
\label{eq:1.9}
\end{equation}
Therefore, the energy of the D-line splitting, $\Delta E_{so}$, can be given by
\begin{equation} 
\Delta E_{so} = E_n^1 \left(j=\frac{3}{2}\right)-E_n^1\left(j=\frac{1}{2}\right)~.
\label{eq:1.10}
\end{equation}
Using Eq. (\ref{eq:1.9}) in this simply leads to
\begin{equation}
\Delta E_{so} =  \frac{Z^2 |E_n| \alpha^2}{2n}~.
\label{eq:1.11}
\end{equation}

\section{Analysis}

The result obtained in the previous section is used here to estimate $\Delta E_{so}$ for the alkali metals from Li to Fr. Obviously, we expect that when applying Eq. (\ref{eq:1.11}) which involves the one electron Bohr energy, the result will not be satisfactory. These calculations were carried out and compared with the experimental values for the D-line splitting energies and listed in Table 1.

\begin{table}[h]
\begin{center}
\caption{\small{D-line Energy Splitting in $10^{-3}$ eV}}

\begin{tabular}{lllll}
\hline \hline
{Atom} & ~~~{Experimental Values}~~~&{References}&~~~{Eq. (\ref{eq:1.10}), Bohr $E_n$}&~~~$n$\\
\hline
Hydrogen& ~~~$4.39023\times10^{-2}$&~\cite{Wayne}&~~~4.528 $\times10^{-2}$&~~~2\\
Lithium& ~~~$4.15641\times10^{-2}$&~\cite{Li Exp}&~~~3.668&~~~2\\
Sodium& ~~~$2.133$&~\cite{Steck Na, Na Exp}&~~~196.0&~~~3\\
Potassium& ~~~$7.159$&~\cite{K Exp}&~~~737.0&~~~4\\
Rubidium& ~~~$29.486$&~\cite{Steck Rb, Rb Exp 1, Rb Exp 2}&~~~5429.0&~~~5\\
Cesium& ~~~$68.756$&~\cite{Steck Cs, Cs Exp 1, Cs Exp 2}&~~~15340&~~~6\\
Francium& ~~~-----&  &~~~60481&~~~7\\
\hline
\end{tabular}
\end{center}
\end{table}

It is clear from these data that when using the Bohr energies in Eq. (\ref{eq:1.9}), with $n$ as listed in Table 1 and $Z$ set to the atomic number of the element, the computed value for $\Delta E_{so}$ significantly overestimates the experimental value for all the alkali metals. 

We propose here that an improved form for Eq. (\ref{eq:1.9}) can be obtained by replacing the Bohr energy, $E_n$, with the first ionization potential for the element involved, which we label as $\Delta E_{ion}$. Experimental values for first ionization potentials for the alkali metals have been tabulated \cite{CRC}. To verify these values, first ionization potentials were also computed using the GAMESS software package \cite{GAMESS} at the DFT/B3LYP level using the MIDI basis set for Li - Cs and the CRENBL effective core potential basis set for Fr \cite{Fr_basis}. These ionization potentials are listed in Table 2.

\begin{table}[h]
\begin{center}
\caption{\small{First ionization potentials for hydrogen and the alkali metals in eV. From experiment and theory.}}

\begin{tabular}{lllll}
\hline \hline
{Atom}&~~~{Exp. \cite{CRC}}&~~~{Computed \cite{GAMESS}}\\
\hline
Hydrogen& ~~~13.59844&~~~13.52349\\
Lithium& ~~~5.39172&~~~5.51992\\
Sodium& ~~~5.13908&~~~5.21234\\
Potassium& ~~~4.34066&~~~4.39092\\
Rubidium& ~~~4.17713&~~~4.15699\\
Cesium& ~~~3.8939&~~~3.76532\\
Francium& ~~~4.0727&~~~1.14500\\
\hline
\end{tabular}
\end{center}
\end{table}

We now used the experimental ionization potentials in place of $E_n$ in Eq. (\ref{eq:1.9}) and re-computed values for $\Delta E_{so}$. These results are listed in Table 3. Obviously, there is a significant improvement over the calculations using the Bohr energy. On comparing these results to the experimental values in Table 1, we see that the best result was for Rb were there is an approximately 3.2\% error between the theoretical result and the known experimental D-line splitting energy.

%\clearpage

\begin{table}[t]
\begin{center}
\caption{\small{D-line Energy Splitting values in $10^{-3}$ eV, from experiment and calculated using Eqs. (\ref{eq:1.9})  and (\ref{eq:1.10}) using experimental ionization potentials from Table 2 in place of Bohr energies. Far right column gives D-line splitting values using the Slater screening constants in Eq. (\ref{eq:1.12}).}}

\begin{tabular}{lllll}
\hline \hline
{Atom} & ~~~{Experimental Values}~~~&{References}&~~~{Eq. (\ref{eq:1.11})}&~~~{Eq. (\ref{eq:1.12}) $S_{sl}$}\\
\hline
Lithium& ~~~$4.15641\times10^{-2}$&~\cite{Li Exp}& ~~~0.646&~~~0.129\\
Sodium& ~~~$2.133$&~\cite{Steck Na, Na Exp}& ~~~5.519&~~~0.314\\
Potassium& ~~~$7.159$&~\cite{K Exp}& ~~~10.43&~~~0.166\\
Rubidium& ~~~$29.486$&~\cite{Steck Rb, Rb Exp 1, Rb Exp 2}& ~~~30.452&~~~1.061\\
Cesium& ~~~$68.756$&~\cite{Steck Cs, Cs Exp 1, Cs Exp 2}& ~~~52.271&~~~0.916\\
Francium& ~~~-----&  &~~~117.315&~~~0.840\\
\hline
\end{tabular}
\end{center}
\end{table}

\section{Discussion}

In the previous section we demonstrated how a
known theory, used to estimate the D-line splitting
energy in the hydrogen atom, can be modified so
as to yield estimates for these values for all of the
alkali metals. The modification was the replacing
of the Bohr energy in the final formula with the
first ionization potential for the element involved.
The energy values computed using this modified
formula were in far greater agreement with the
experimental values than those computed when
using the Bohr energy. Clearly, the use of the
ionization potential in the place of the Bohr energy
helps correct the result by implicitly including
screening effects in the multi-electron atoms.

Upon comparing our results for the D line splitting energy in Table 3 with the experimen-
tal values it can be seen that this scheme still overestimate the values for the elements lighter
than Rb but underestimates the known value for Cs and likely also for Fr. That is, using an
ionization potential to replace the Bohr energy in Eq. (\ref{eq:1.9}) underestimates screening for the
lighter atoms but over estimates it for the heavier. We suspect that second order corrections
to the energy will be too small to account for the remaining discrepancy and rather seek to
find a corrected screening model for the Bohr energy that might be introduced in Eq. (\ref{eq:1.9}).

Many semi-empirical schemes have been developed for using the hydrogenic model to estimate total energies, polarizabilities and ionization potentials for multi-electron atoms by replacing $Z$ by an effective $Z$, $Z_{eff}$, where $Z_{eff} = Z-S$ \cite{Slater, Lang, Agmon, Bird}. Here $S$ is a screening constant. 
Also, the principle quantum number can be replaced with an effective principle quantum number $n^{*}$. 

A well-known method for computing the
hydrogenic screening constant was developed, in
the early twentieth century, by J. C. Slater \cite{Slater}.
Using this prescription, we computed the screening
constant for the outer $s$ electron in the group one
atoms and labeled these as $S_{sl}$ . These screening
constant values are listed in Table 3. When
directly using these values in $Z_{eff}$ , along with the
corresponding values for $n^*$ listed in Reference
\cite{Bird}, in Eq. (9), we find that the values for $\Delta E_{so}$ are off, often by an order of magnitude. These
values are also listed in Table 3.

The Slater scheme can be used to generate accurate values for first ionization potentials, most notably, for elements in the first four rows of the
periodic table. This is accomplished by finding
the screening constant for the outer electron in the
neutral atom and then again for the outer electron
in the cation. Then, the total energy for each can
be computed and the difference between these
found. However, since the experimental values for
the first ionization potentials, when used in Eq.
(\ref{eq:1.11}) in place of the Bohr energy, fail to yield
accurate values for the D-line splitting energy for
all of the alkali metals it is no surprise that Slater
ionization potentials, nor the Slater hydrogenic
screening constants, would serve to yield accurate
estimates for $\Delta E_{so}$ either. We therefore seek a
screening constant that would be effective when
using Eq. (\ref{eq:1.11}) to compute the D-line splitting
energies in the alkali metals.

First we use the experimental data for $\Delta E_{so}$ to find
an empirical relation for S. It is a straight forward
thing to use Eq. (\ref{eq:1.11}), and the known values for
$\Delta E_{so}$ , to compute an estimate for S for each of the
elements considered here. Eq. (\ref{eq:1.11}) adjusted to
allowing for a screening constant is simply
\begin{equation}
\Delta E_{so} =  \frac{(Z-S)^4 (13.6) \alpha^2}{2(n^*)^3}~~~\text{eV}~,
\label{eq:1.12}
\end{equation}
where we let $n^*$ take on the Slater values \cite{Bird} and
$\Delta E_{so}$ has units of electron-volt (eV). Solving this for $S$ we get
\begin{equation}
S = Z - \left[\frac{2(n^* )^3 \Delta E_{so}}{(13.6) \alpha^2}\right]^{1/4}~,
\label{eq:1.13}
\end{equation}
Using the experimental values for $\Delta E_{so}$ , screening
constants for the alkali metals are computed using
Eq. (\ref{eq:1.13}) and listed in Table 4 along with those
generated using the Slater scheme.

\begin{table}[t]
\begin{center}
\caption{\small{Screening constants, $S$, computed by using Eq. (\ref{eq:1.13}) and $S_{sl}$ computed using the Slater
scheme for the outer $s$ electron. Effective principle quantum numbers were taken from \cite{Bird}
except in the case of Francium.}}

\begin{tabular}{lllll}
\hline \hline
{Atom} & ~~~{$S$ Eq. (\ref{eq:1.14})}~~~&{$S_{sl}$ \cite{Bird}}&~~~{$Z$}&~~~$n^*$\cite{Bird}\\
\hline
Lithium& ~~~2.02&~~~1.7&~~~ 3&~~~ 2\\
Sodium& ~~~7.45&~~~8.8&~~~ 11&~~~ 3\\
Potassium& ~~~13.37&~~~16.8&~~~ 19&~~~ 3.7\\
Rubidium& ~~~28.50&~~~33.3&~~~ 37&~~~ 4.0\\
Cesium& ~~~44.11&~~~51.3&~~~ 55&~~~ 4.2\\
Francium& ~~~76.16& ~~~83.3 &~~~ 87&~~~ 4.3$^1$\\
\hline
\end{tabular}
\end{center}
\end{table}

Obviously, the screening constants computed with Eq. (\ref{eq:1.13}) are a function of $Z$. A plot of these versus $Z$ is shown in Figure 2. We find that $S(Z)$ obeys a quadratic law given by
\begin{equation}
S = a + bZ + cZ^2~.
\label{eq:1.14}
\end{equation}
A curve of this type is fit to these data and shown in Figure 2. The fit was excellent having an R-squared value of 0.999. 

%\clearpage
\begin{figure}[h]
  \centering
  %\label{fig:7}
  \includegraphics[width = 8.0 cm]{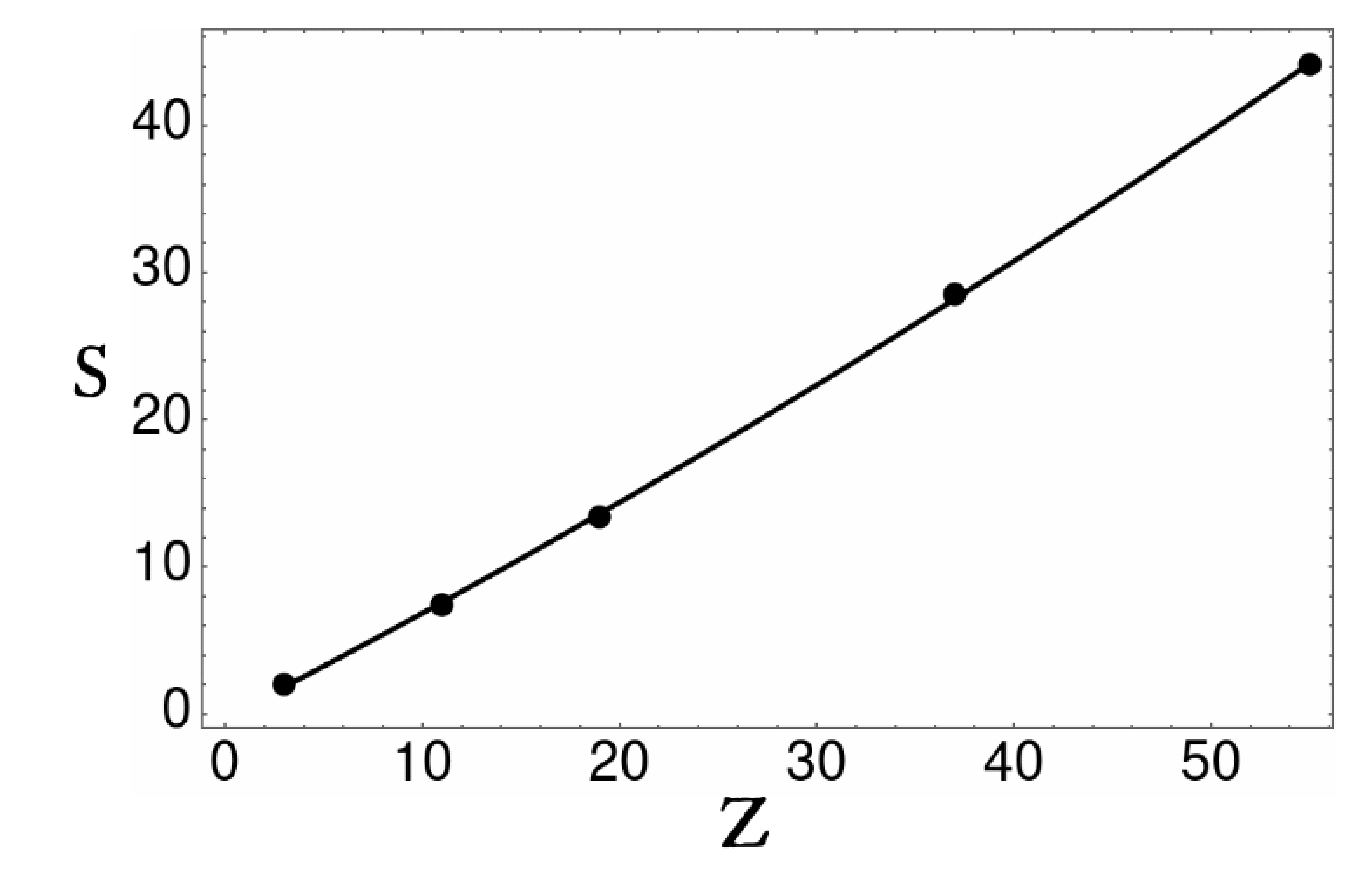}
  \caption{Plot of $S$ values, computed using Eq. (\ref{eq:1.13}), vs. $Z$. The fit parameters from Eq. (\ref{eq:1.14}) were $a$ = -0.235754, 
  $b$ = 0.686758 and $c$ = 0.00220759.}
  %\label{fig1}
\end{figure}

We develop a simple model for $S$ by considering the classical model for the electric field magnitude, $\mathcal{E}$ at the outer $s$ electron. 
Consider a Gaussian spherical surface of radius $r$ which encloses the charge of the nucleus, $Ze$ and some fraction of the charge, $-(Z-1)e$. due to the 
electron cloud excluding the outer electron. Quantum mechanics predicts that the inner electrons may spend part of the time at 
greater distance from the nucleus that the outer $s$ electron so that, when averaged over time, some fraction of the electron cloud is enclosed within the Gaussian 
surface at the location $r$, of the outer $s$ electron. In Gauss's law this gives
\begin{equation}
\mathcal{E} = \frac{Ze - (Z-1)e \frac{r^3}{R^3}}{4\pi \epsilon_o r^2}~,
\label{eq:1.15}
\end{equation}
where $r < R$. With a little manipulation, one finds that the numerator of the right side of Eq. (\ref{eq:1.15}) can be written as
\begin{equation}
e\left[Z - \left(Z\frac{r^3}{R^3} - \frac{r^3}{R^3}\right) \right]~,
\label{eq:1.16}
\end{equation}
so that we identify the screening constant as
\begin{equation}
S = \left(\frac{r}{R}\right)^3 (Z - 1)~.
\label{eq:1.17}
\end{equation}

The exact value to use for the radius $r$ is unknown but, we expect that it will vary with $Z$ due to relativistic effects. It is known that the velocity of inner shell electrons, being directly proportional to $Z$, can approach a sizable fraction of the speed of light which in turn leads to mass enhancement which contracts the orbital and causes a higher screening of the nuclear charge as $Z$ increases \cite{Blaha}. Therefore we propose that, due to such effects effects, the value for $r$ contracts on going down the group one column. This can be modeled by an expression of the form \cite{Ter, Trago}.

\begin{equation}
r = r_o \sqrt{1 - \left(\frac{Z}{nc}\right)^2 }~.
\label{eq:1.18}
\end{equation}
Here $n$ is a quantum number and $c$ the speed of light. Clearly $Z/(nc) << 1$ so that on cubing both sides of Eq. (\ref{eq:1.18}) it can be expanded to arbitrary order in a Maclaurin series as
\begin{equation}
r^3 = {r_o}^3 \left(1 - \frac{3}{2}\left(\frac{Z}{nc}\right)^2 + \frac{3}{8}\left(\frac{Z}{nc}\right)^4 + \frac{1}{16}\left(\frac{Z}{nc}\right)^6 + \frac{3}{128}\left(\frac{Z}{nc}\right)^8 + \mathcal{O}^8 \right)~.
\label{eq:1.19}
\end{equation}
Using Eq. (\ref{eq:1.19}) for $r$ in Eq. (\ref{eq:1.17}), then rearranging and factoring yields
\begin{equation}
S = \left(\frac{r_o}{R}\right)^3 \left[Z - 1 +Z^2 \left(\frac{3}{2(nc)^2} - \frac{3Z}{2(nc)^2}-\frac{3Z^2}{8(nc)^4}+\frac{3Z^3}{8(nc)^4}-\frac{Z^4}{16(nc)^6}+ \mathcal{O}^7 \right)\right]~.
\label{eq:1.20}
\end{equation}
Now, in this case $3 \le Z \le 87$ and as $nc >> Z$ the terms in parenthesis of Eq. (\ref{eq:1.20}) are certainly small and getting progressively smaller so that we set the sequence in parenthesis to be the constant $\gamma$. Therefore, the model gives as a final expression for the screening constant
\begin{equation}
S = -\left(\frac{r_o}{R}\right)^3 + \left(\frac{r_o}{R}\right)^3 Z + \left(\frac{r_o}{R}\right)^3 \gamma Z^2~,
\label{eq:1.21}
\end{equation}
which is in agreement with the empirical result given in Eq. (\ref{eq:1.14}).

The precise value to use for the constant $r_o$ is not
known to the authors. However, as a first estimate
for the ratio $r_o /R$ we let $r_o$ be the cation radius and,
as mentioned previously, $R$ is the covalent radius.
From tabulated data \cite{Davis}, one finds that $(r_o /R)^3$
ranges from 0.208 for Li to 0.319 for Cs. On
comparing Eqs. (\ref{eq:1.14}) and (\ref{eq:1.21}) we have $a = b = (r o /R)^3$.
The values for the constants $a$ and $b$ from the
curve fit of Figure 2 are $a = 0.236$ and $b = 0.687$.
As the agreement between $(r_o /R)^3$ and $a$ is excellent
for Li and, within an order of magnitude for $b$,
one must assume we are on the right track and we
therefore suggest that the cube of the cation to
covalent radius would be a reasonable choice for
$a$ and $b$ when using Eq. (\ref{eq:1.21}) to estimate a
screening constant. Further research is necessary
to clarify the exact nature of $r_o$.

On examining Table 4, one finds that the Slater screening constants underestimate the screening for Li and overestimate screening for the other alkali metals. This was the same trend mentioned earlier when considering the effect of using ionization potentials to compute D-line splitting energies which are listed in Table 3. This difference in the intensity of screening between the two models can be attributed to the electron-relaxation effect \cite{Lang} whereby during ionization the atom contracts and thus slightly lowers the required energy for ionization from what one might expect when using a rigid, time-independent model of the atom. Therefore, the screening constants used to predict ionization energies implicitly correct for this effect by inflating the screening constant. However, electron relaxation is not an effect one must account for when modeling the spin-orbit phenomenon and therefore the screening constants in this case need not be overestimated.

\section{Conclusion}

In this report we demonstrated how a well known result for the D-line splitting energy in hydrogen can be modified and then used to estimate this value for all of the alkali metals. The alteration is in adjusting the Bohr energy with an appropriate screening constant which is a function of the atomic number. Through an empirical fit the screening constant is shown to be a mildly quadratic function of the atomic number. Using a simple model, we show how this form of the screening constant can be derived from first principles with a relativistic correction. It is hoped that this result might find use in predicting fine structure splitting energies for other atomic groups from the periodic table.

\section{Acknowledgements}

The authors wish to thank the College of Arts and Sciences at The University of the South for funding support.

\end{document}